\documentclass[%
 twocolumn,
superscriptaddress,
 notitlepage,
 showpacs,
 amsmath,amssymb,
 aps,
prb,
floatfix
]{revtex4-1}

\usepackage{graphicx}
\usepackage{dcolumn}
\usepackage{bm}
\usepackage{xcolor}
\usepackage{braket}
\usepackage{hyperref}

\begin{document}

\title{Many-Body Localization in Translational Invariant Diamond Ladders with
  Flat Bands}
\author{Mirko Daumann}
\affiliation{
 Fakult\"at f\"ur Physik, Universit\"at Bielefeld, Postfach 100131, D-33501 Bielefeld, Germany
}
\author{Robin Steinigeweg}
\affiliation{
 Fachbereich Mathematik/Informatik/Physik, Universit\"at Osnabr\"uck, D-49076 
Osnabr\"uck
}
\author{Thomas Dahm}
 \email{thomas.dahm@uni-bielefeld.de}
\affiliation{
 Fakult\"at f\"ur Physik, Universit\"at Bielefeld, Postfach 100131, D-33501 Bielefeld, Germany
}

\date{\today}

\begin{abstract}
The presence of flat bands is a source of localization in lattice systems. 
While flat bands are often unstable with respect to interactions between the particles, they can persist in certain cases.
We consider a diamond ladder with transverse hopping that possesses such stable flat bands and show that many-body localization appears in the presence of interactions without quenched disorder.
We numerically demonstrate that the eigenstate thermalization hypothesis is violated. We show that the influence of degenerate flat band states spreads across the full energy spectrum and grows extensively in the thermodynamic limit.
Furthermore, we verify localization in terms of time evolution of some local observable, revival probability, participation ratio and entanglement entropy.
\end{abstract}
\maketitle
\section{Introduction}
\label{sec:intro}
\begin{figure}
	\centering
	\includegraphics[width=\linewidth]{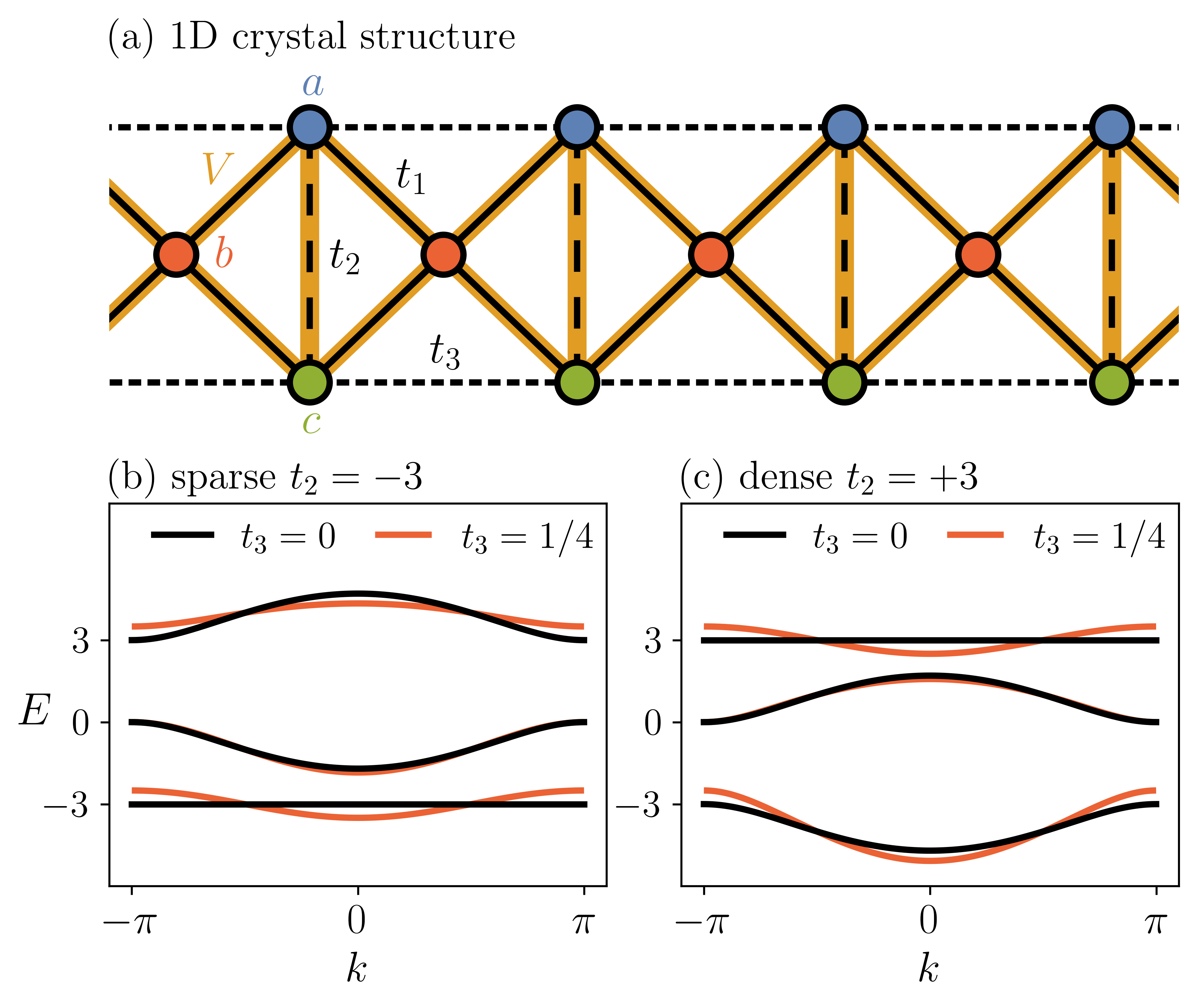}
	\caption{Quasi one-dimensional lattice (diamond ladder) with periodic boundaries containing FBs. 
	(a) Crystal structure with three orbitals $a$, $b$, $c$ per unit cell and three kinds of hoppings $t_1$ [solid], $t_2$ [dashed] and $t_3$ [dotted].
	$V$ represents particle-particle interaction along solid and dashed lines.
	(b) Single-particle band structure [black] as a function of dimensionless momentum $k$ for the sparse system ($t_2=-3$). 
	The FB lies below the other two bands and may be made dispersive for $t_3\neq0$ [red]. (c) Same for the dense system ($t_2=+3$). 
	The FB lies above the other bands.}
	\label{fig:model}
\end{figure}

Many-body localization (MBL) is usually seen as a generalization of Anderson localization \cite{1958Anderson} to interacting systems and thus is discussed to occur in disordered systems \cite{1980Fleishman,1997Altshuler,2005Gornyi,2006Basko,2007Oganesyan,2008Znidaric,2010Pal,2015Altman,2015Nandkishore,2015Schreiber,2016Imbrie,2017Abanin,2018Alet,2019Abanin}.
These systems do not thermalize even in the presence of an interaction between the particles. 
Systems that exhibit MBL have been studied both theoretically and experimentally, because they can retain some memory of the initial conditions and are thus of interest for storing quantum information \cite{2015Nandkishore,2015Schreiber,2019Abanin} and hence may implement quantum memory devices. 
However, disorder breaks the translational invariance of these systems.
There has been considerable discussion whether MBL could also exist in systems without disorder \cite{2012Carleo,2014Grover,2014Roeck1,2014Roeck2,2014Schiulaz,2015Schiulaz,2015Papic,2015Horssen,2016He,2016Hickey,2016Kim,2016Pino,2016Yao,2017Mondaini,2017Prem,2017Sierant,2017Smith1,2017Smith2,2018Brenes,2018Lan,2018Smith,2018Yarloo,2019Lerose,2019Nieuwenburg,2019Schulz,2019Sirker,2020Heitmann1,2020Khare,2020Khemani,2020Rakovszky,2020Sala,2021Karpov,2021Scherg,2021Hart,2022Zisling}.
Recently, flat band (FB) systems have been considered in this context \cite{2020Kuno,2020Roy,2021Orito,2020Danieli,2020Li,2020McClarty,2021Vakulchyk,2022Danieli}.
Noninteracting FB systems are characterized by the presence of flat energy dispersions and thus possess highly degenerate energy eigenstates. 
This in turn allows stationary, strongly localized states in the system, also known as compact localized states\cite{2015Derzhko,2017Maimaiti,2018Leykam}.
As a result, the FB is a different source of localization in these systems and could equally well (instead of Anderson localization) lead to interacting many-body quantum systems that possess localization in the presence of particle-particle interactions.
FBs are often unstable against interaction, though \cite{2018Leykam}.
However, there exist specific models where FBs can survive in presence of interaction due to symmetry protection \cite{2020Tilleke} or in detangled systems \cite{2020Danieli,2022Danieli}, for example.
Here, we study the one-dimensional diamond ladder [Fig.~\ref{fig:model}(a)] for spinless fermions, which previously was shown to host such stable FBs \cite{2020Tilleke}.
The question of thermalization is closely related to the so-called eigenstate thermalization hypothesis (ETH) \cite{1991Deutsch,1994Srednicki,2008Rigol}, which postulates that every energy eigenstate yields expectation values close to the corresponding microcanonical values. 
MBL systems have been found to violate ETH, see e.g. \cite{2015Nandkishore,2010Pal}. 
We will provide evidence for MBL in the one-dimensional diamond ladder by showing that the ETH is violated in this system. 
We support our results by an analysis on how degenerate FB states influence the spectrum at almost any energy. 
Without interaction the fraction of FB-influenced states dominates in the thermodynamic limit and still grows extensively in presence of interaction.
The results are backed up by time evolutions for some local observable, revival probability, participation ratio and entanglement entropy (EE) for nonequilibrium initial states, quantities that characterize localization in many-body quantum systems \cite{2015Beugeling,2017Santos,2020Santos}. 
Scaling with system size will be discussed throughout this paper.
\section{Model}
\label{sec:mmodel}
The Hamiltonian of the system reads:
\begin{equation}
	H=-\sum_{\langle i\alpha,j\beta\rangle} t_{i\alpha,j\beta}\cdot c^\dagger_{j\beta}c_{i\alpha}+\frac{V}{2}\sum_{\langle\langle i\alpha,j\beta\rangle\rangle} n_{j\beta}n_{i\alpha}\ ,
\end{equation}
with fermionic creation (annihilation) operators $c^\dagger_{i\alpha}$ ($c_{i\alpha}$) in unit cell $i$ for the orbital $\alpha\in\left\{a,b,c\right\}$ and particle density operators $n_{i\alpha}=c^\dagger_{i\alpha}c_{i\alpha}$.
$t_{i a,j b}=t_{i c,j b}=t_1$ describes nearest-neighbor hopping, $t_{i a,i c}=t_2$ stands for next-nearest-neighbor hopping within same unit cells, and $t_{i a,i\pm1a}=t_{i c,i\pm1c}=t_3$ permits hopping to next-nearest neighbors of adjacent cells as illustrated in Fig.~\ref{fig:model}(a). 
In the following we will choose $t_1$ as the unit of energy and thus set $t_1=1$.
$V$ indicates (next-)nearest-neighbor repulsive particle-particle interaction.
$\langle i\alpha,j\beta\rangle$ stands for a summation over all nearest- and next-nearest-neighbored lattice sites $i$ and $j$ with orbitals $\alpha$ and $\beta$ along all types of lines in Fig.~\ref{fig:model}(a). 
$\langle\langle i\alpha,j\beta\rangle\rangle$ omits summation over dotted lines. Thus, the interaction only acts on the original diamond ladder structure.
The transversal hopping amplitude $t_2=\pm3$ determines whether the FB lies below or above the other bands [see Fig.~\ref{fig:model}(b) and (c)].
Throughout this work we focus on the case where the FB is half filled in the groundstate. 
For $t_2=-3$ the system is thus $1/6$-filled and we denote this as sparse system, while for $t_2=+3$ on the other hand we have a filling of $5/6$ which is denoted as dense system. 
We introduce a perturbative inter-cell hopping $t_3$ to purposely turn on a dispersion of the FB in order to compare our results between presence and absence of a FB.
Below, we will discuss the case $t_3=1/4$ [see red curves in Fig.~\ref{fig:model}(b) and (c)] as a comparison for a system without FB. 
In a previous work \cite{2020Tilleke} we have shown that FB groundstates in this system are robust in terms of properties like degeneracy when turning on particle-particle interaction.
This holds for interaction values below $V\approx1.8$ in the case of a dense population, yet for unconditionally large $V$ values in the sparse case.
\section{Eigenstate Thermalization Hypothesis}
\label{sec:eth}
As a first step to assess whether this interacting system shows MBL, we numerically test if the ETH is broken or not. 
In a system that thermalizes and fulfills the ETH one expects that expectation values of observables $\braket{A}$ in energy eigenstates with similar energy do not fluctuate much from one eigenstate to another.
If a system possesses stationary and localized states, however, we expect the ETH to be broken.
Then, the states possess local information which leads to an higher amount of fluctuations of $\braket{A}$. 
In the thermodynamic limit these fluctuations will remain finite, while they will go to zero, when ETH is fulfilled.

\begin{figure}
	\centering
	\includegraphics[width=\linewidth]{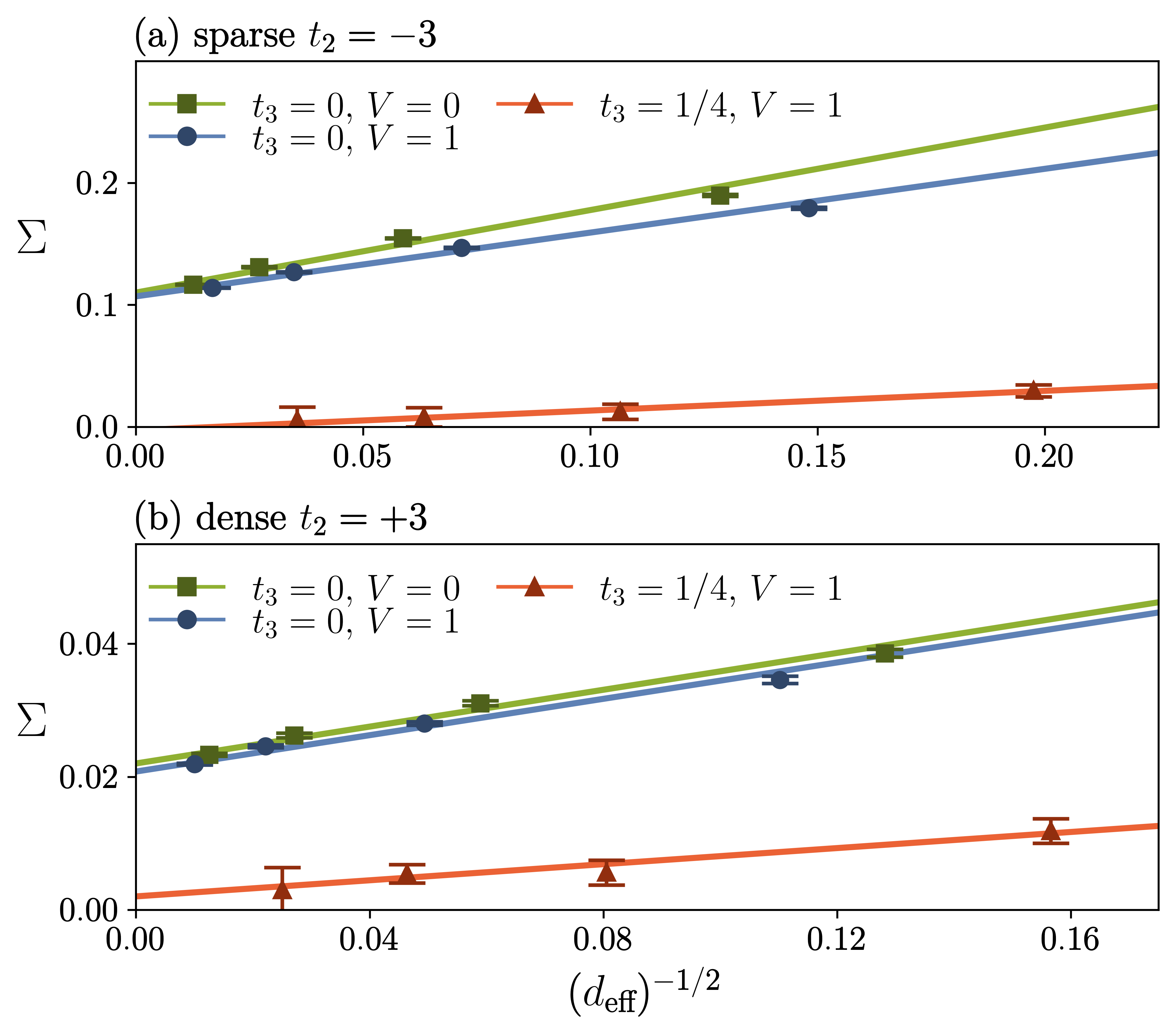}
	\caption{Results of the ETH calculations.
	(a) Decrease of expectation value fluctuations with increasing system size for the sparse system. 
	$\Sigma$ only tends to zero if $t_3\neq 0$ and the FB becomes dispersive.
	(b) Corresponding results for the dense system.
	Error bars show the statistical error coming from the average over a finite number of random states $\ket{\Psi}$.}
	\label{fig:eth}
\end{figure}
To test ETH numerically, we use the method of Ref.~\cite{2013Steinigeweg} which is based on an accurate scheme using typical pure states drawn at random \footnote{For other applications of random states see   T. Heitmann, J. Richter, D. Schubert, and R. Steinigeweg, Z. Naturforsch. A. {\bf 75}, 421 (2020).}.
An energy filter around energy $U$ with width $\sigma$ is applied on a Gaussian distributed (mean zero) normalized random state $\ket{\Psi}$:
\begin{equation}
	\ket{\Phi}=\exp\left[-\frac{\left(H-U\right)^2}{4\sigma^2}\right]\cdot \ket{\Psi}\ .
	\label{eq:filter}
\end{equation}
We chose $U$ to be equal to the ground-state energy and a small $\sigma=3/2$.
$\ket{\Phi}$ then contains groundstates and a mix of energetically low-lying excited states. 
The quantity
\begin{equation}
	\Sigma=\sqrt{\frac{\left\{\int_t\braket{\Phi|A(t)A|\Phi}\right\}}{\left\{\braket{\Phi|\Phi}\right\}}-\frac{\left\{\braket{\Phi|A|\Phi}\right\}^2}{\left\{\braket{\Phi|\Phi}\right\}^2}}\ 
	\label{eq:Sigma}
\end{equation}
measures fluctuations of the expectation values of $A$.
Here, the curly brackets indicate that these values are averaged over different random states $\ket{\Psi}$.
The time average is $\int_t=1/(t_\text{e}-t_\text{i})\int_{t_\text{i}}^{t_\text{e}}\text{d}t$ where $t_\text{i}$ and $t_\text{e}$ are suitably chosen such that the integral best matches the relaxation value of $\braket{\Phi|A(t)A|\Phi}$ \cite{2013Steinigeweg}.
The first fraction therefore serves as long-term auto correlation.
The second term is simply the squared expectation value of $A$ in the filtered state.
Its square root is a good estimate of the actual microcanonical ensemble average at energy $U$.
Deviations between the first and second term will typically become smaller when the
system size is increased.
They will go to zero if the states thermalize.
However, if the system breaks ergodicity and preserves information of the initial state over long times, the fluctuations will not vanish and $\Sigma$ will take a finite value.

For the operator $A$ we choose:
\begin{equation}
	A=\frac{N}{M\left(N-1\right)}\sum_{i\alpha}\sin^2\left(\frac{\left(i-1\right)\cdot\pi}{N-1}\right)\cdot n_{i\alpha}\ .
	\label{eq:A}
\end{equation}
$N$ is the number of unit cells and $M$ is the number of particles.
This quantity detects particle density inhomogeneities along the lattice. 
For homogeneously distributed states the expectation value is $1/2$.

In Fig.~\ref{fig:eth} we present our results for the sparse and dense system with different system sizes.
Each regression line contains four points for $N=6, 8, 10, 12$ unit cells and $M=3,4,5,6$ particles in the sparse and $M=15,20,25,30$ particles in the dense setting. 
The Hilbert space dimension is $d=\binom{3\cdot N}{M}$ and ranges from 816 to 1.9 million.
The effective dimension $d_\text{eff}$ that is probed by the state $\ket{\Phi}$ is calculated from:
\begin{equation}
	d_\text{eff}=d\cdot\left\{\braket{\Phi|\Phi}\right\}\ .
\end{equation}
Matrix exponentials and time evolutions in (\ref{eq:filter}) \& (\ref{eq:Sigma}) and in the following sections are performed via Krylov subspace algorithms \cite{1950Lanczos,2003Moler,2006Mohankumar}.
We have chosen the interaction strength $V=1$ in order to stay below the critical value of $\approx 1.8$ \cite{2020Tilleke}.
For the dispersive system ($t_3=1/4$) the fluctuations scale with the inverse square root of the effective dimension, which one should expect for generic nonintegrable systems \cite{1999Srednicki,2014Beugeling}.
In contrast, if the FB is present ($t_3=0$), the fluctuations do not vanish in the limit $d_\text{eff} \rightarrow \infty$ both with and without interactions.
This observation holds for both the sparse and the dense system, while in the latter case the fluctuations are smaller roughly by a factor $1/5$ due to the $5$-fold filling compared to the sparse system which enters in the definition (\ref{eq:A}) of operator $A$.
Only if we turn on the dispersion via $t_3$, the system shows thermalization.
This is one of our central results, as it shows that ETH remains violated also in the presence of interaction.
\section{Scaling of Degeneracy}
\label{sec:degen}
The results above show a breakdown of the ETH for states near the groundstate energy. 
\begin{figure}
	\centering
	\includegraphics[width=\linewidth]{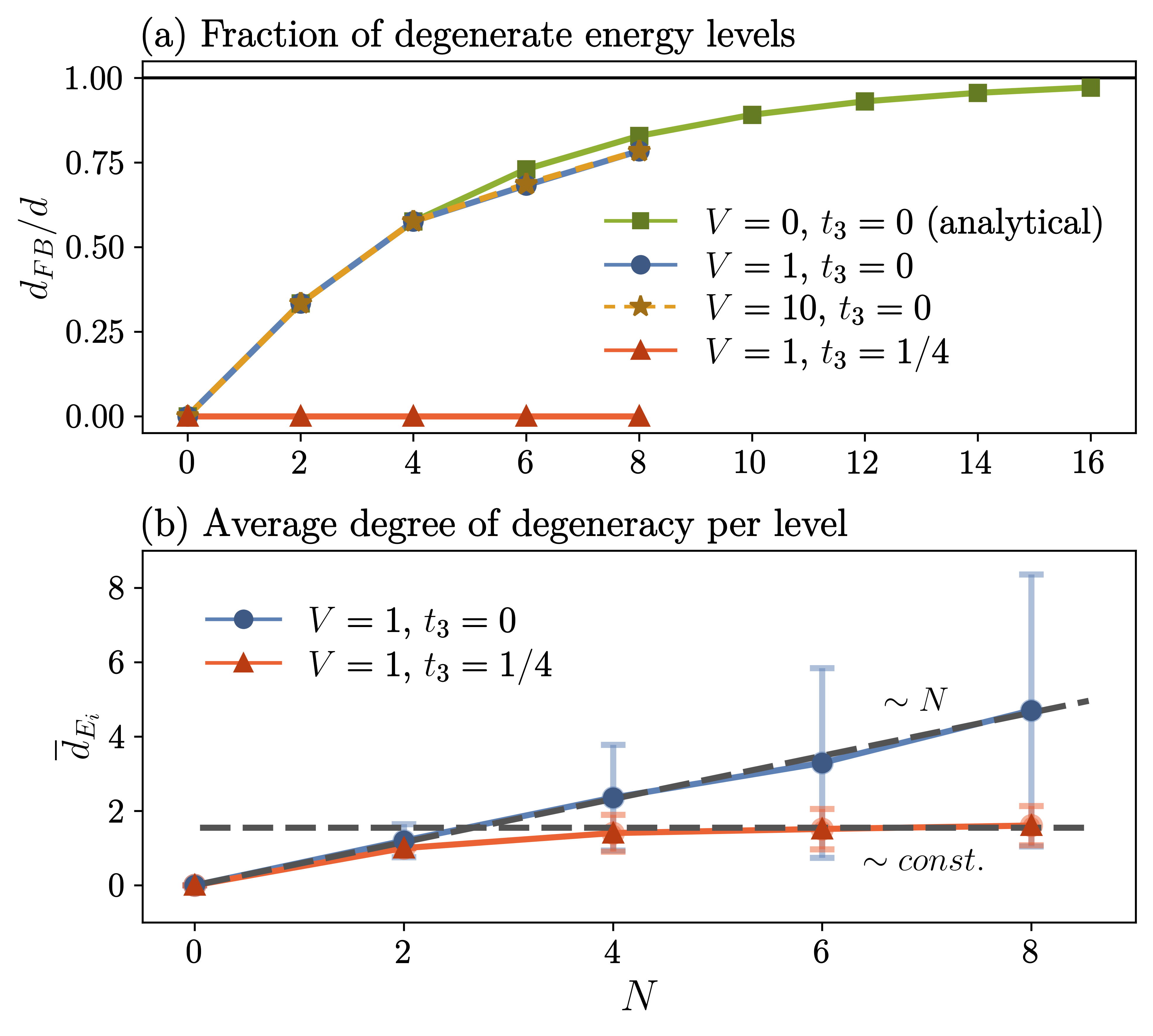}
	\caption{Degeneracy data of a $1/6$ filled system with increasing system size.
	(a) Fraction of degenerate energy levels $d_{FB}$ asymptotically approaches the full Hilbert space dimension $d$ in the thermodynamic limit.
	This trend is followed by numerical data for $V=1$ and $V=10$ in accessible system sizes while no significant amount of degenerate levels can be detected in absence of the FB [red curve].
	(b) Average degree of degeneracy $\overline{d}_{e_i}$ of every energy level $d_{e_i}$ with standard deviation as error bars. 
	When the FB is unperturbed $\overline{d}_{e_i}$ grows extensively with system size, while it becomes constant in the perturbed case.}
	\label{fig:degen}
\end{figure}
ETH can be checked across the whole spectrum in a similar manner by adjusting the filter energy $U$ for higher energies.
As shown in Appendix A we observe the ETH to be violated almost across the whole spectrum.
Here we additionally want to present an exact calculus for the $V=0$ case which gives an expression how numerous FB states are in the full Hilbert space with increasing system size based on their degeneracy.
Numerical results from the scenario with interaction $V=1$ and $V=10$ are compared to the $V=0$ results.
	
Degenerate energy levels are highly relevant in terms of localization.
They allow reorganization of eigenstates in the corresponding eigenspace which then can lead to the existence of a spatially localized structure in these subspaces.
Extensive scaling of degeneracy therefore can serve as a reliable predictor whether localization due to FBs survives when going to the thermodynamic limit.
Without interaction, any eigenstate $\ket{\varphi_{m_1,m_2,m_3}(\{k_i\})}$ for $m_1+m_2+m_3=M$ particles can be written as product state in terms of single-particle band excitations $c^\dagger_{\alpha_i}(k_i)$ which each adds a particle with band index $\alpha_i=1,2,3$ and momentum $k_i$ to the system:
\begin{equation}
	\ket{\varphi_{m_1,m_2,m_3}(\{k_i\})}=c^\dagger_{\alpha_1}(k_1)\dots c^\dagger_{\alpha_M}(k_M)\ket{0}\ .
	\label{eq:phiproduct}
\end{equation}
$m_1$ gives the total number of particles in the FB and $m_2$ \& $m_3$ in either of the other two dispersive bands.
	
Due to the flat energy dispersion of the FB, any level in the Hilbert space will bear not less than an $N$-fold degeneracy if $m_1>0$.
The total dimension of degenerate basis states caused by the FB is given by:
\begin{equation}
	d_{FB}=\sum_{m_1=1}^{M}\binom{N}{m_1}\sum_{m_2=0}^{M-m_2}\binom{N}{m_2}\binom{N}{\underbrace{M-m_1-m_2}_{m_3}}\ .
\end{equation}
This formula also gives correct results for $m_i>N$ with $\binom{N}{m_i}=0$ in that case. 
These considerations work similarly regarding $3\cdot N-M$ holes instead of $M$ particles for a densely filled system with $M>3\cdot N/2$.

The fraction of $d_{FB}$ divided by the total Hilbert space dimension $d$ quickly approaches $1$ for a fixed particle density in the thermodynamic limit meaning that asymptotically the whole spectrum becomes degenerate due to the FB in large systems.
This is shown as green curve for a filling of $1/6$ in Fig.~\ref{fig:degen}(a).
The same can be tested numerically for small systems [blue, yellow and red curve].
In order to rule out non-FB related degeneracy - e.g. from symmetric points in the spectrum - when numerically calculating $d_{FB}$, only degeneracy multiples are considered which can be tied to the FB.
This means only if the degeneracy of some energy level $d_{E_i}$ divided by any $\binom{N}{m_1}$ (with $m_1=1\dots M$) is without remainder it is taken into account.
Given the course of the blue and yellow curve in Fig.~\ref{fig:degen}(a) we can conclude that presence of interaction, even for higher values like $V=10$, does not change the overall trend of an extensive growth of degeneracy.
	
This is underpinned by the linear scaling of the average degree of degeneracy $\overline{d}_{E_i}$ with system size $N$ as blue curve shown in Fig.~\ref{fig:degen}(b).
$\overline{d}_{E_i}$ is calculated as arithmetic mean of all $d_{E_i}$ with standard deviation as error bars.
The error bars are growing simply due to the fact that occurring $d_{E_i}$ are subjected to an increasing variance between $\binom{N}{1}$ and $\binom{N}{M}$.
In contrast, both red curves in Fig.~\ref{fig:degen} for a perturbed FB neither show a noticeable appearance of total FB related degeneracy $d_{FB}$, nor an extensive scaling of $\overline{d}_{E_i}$.
\section{Dynamics of Typical Nonequilibrium States}
\label{sec:time}
Next, we take a look at the time evolution of nonequilibrium states and examine the influence of the FB.
\begin{figure}
	\centering
	\includegraphics[width=\linewidth]{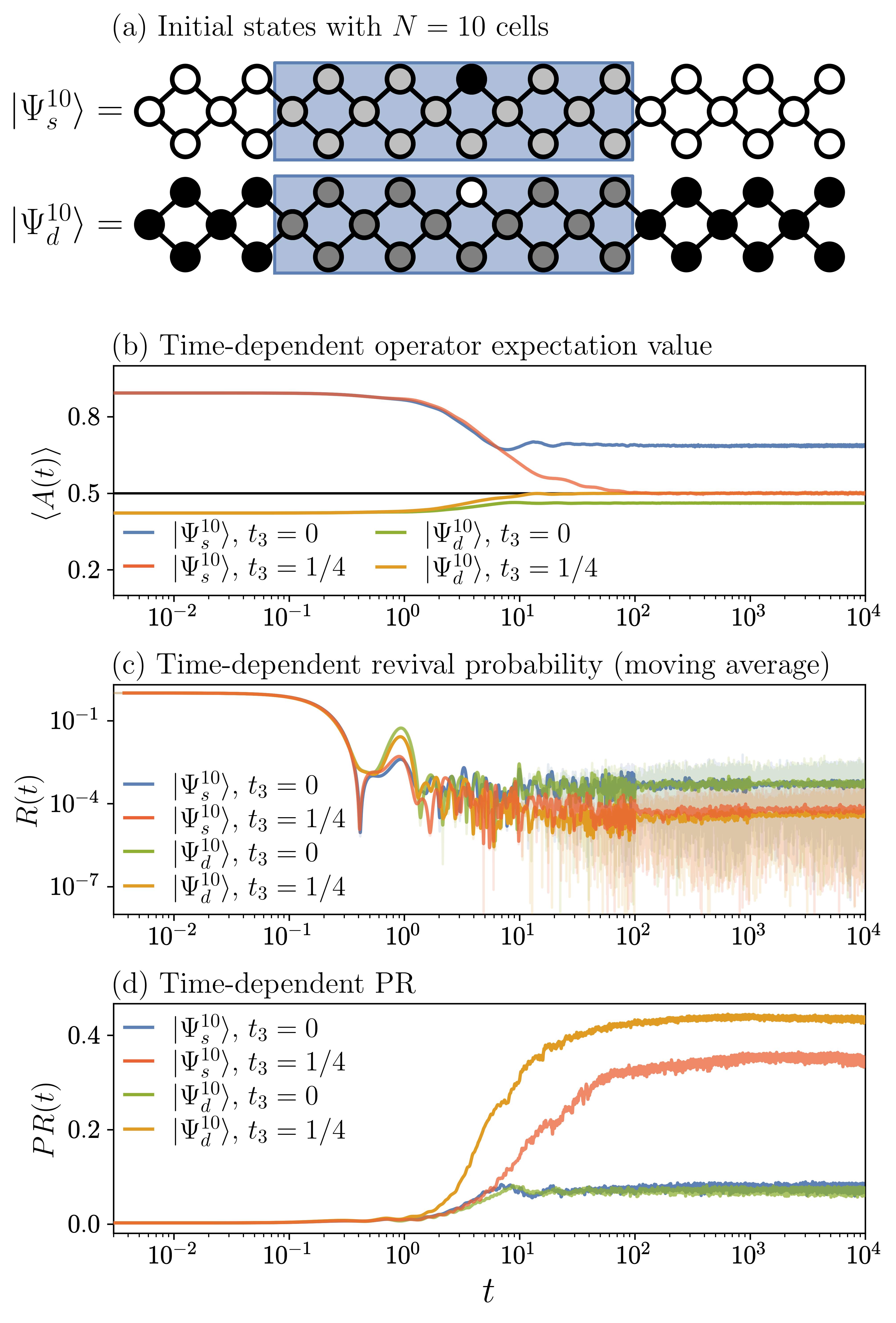}
	\caption{Time evolution with interaction $V=1$ for a system with 10 unit cells on a logarithmic time scale in units of inverse energy $1/t_1=1$. 
	The data are calculated for single initial states.
	(a) $\ket{\Psi_\text{s}^{10}}$ and $\ket{\Psi_\text{d}^{10}}$ in real space representation are the initial states for the sparse and dense system, respectively. 
	(b) The evolution of inhomogeneity in terms of the operator $A$. 
	(c) Probability for the evolved state to return into its initial condition. 
	Due to strong fluctuations, we show the moving average (intense color) above the original data (light color). 
	(d) Evolution of the participation ratio as measure for localization.}
	\label{fig:time}
\end{figure}
In real space representation, the initial states [~\ref{fig:time}(a)] combine a strongly localized density peak [black dot, sparse] or dip [white dot, dense] with a randomly filled background in $N'$ cells [gray dots in blue shaded area]. 
In this central area, a particle/hole is guaranteed to be found at the middle $a$ site whereas the other particles/holes are distributed over neighbored cells.
The central region is bordered by an homogeneous surrounding of empty or filled cells for the sparse and dense case respectively. 
This surrounding can be expanded at will.
	
Generating the central area is identical to previous works in Ref.~\cite{2014Karrasch,2017Steinigeweg,2020Heitmann1}:
\begin{equation}
	\ket{\varphi_{\text{s,d}}^{N'}}\sim\left\{ 
	\begin{array}{r}
		n_{N'/2,a}\cdot\ket{\Psi} \ \text{sparse}\\
		\left(1-n_{N'/2,a}\right)\cdot\ket{\Psi} \ \text{dense }
	\end{array}\right.\ ,
\end{equation}
where $n_{N'/2,a}$ is the particle number operator for the position of the peak and $\ket{\Psi}$ is, as before, a Gaussian distributed random vector in particle number eigenbasis.
In this work, the central area always consists of $N'=5$ cells containing five particles or five holes.
After generation of this region, the coefficients of $\ket{\varphi_{\text{s,d}}^{N'}}$ can be embedded in larger Hilbert spaces $\binom{3\cdot N}{5}$ with empty cells outside the center for states $\ket{\Psi_s^{N}}$ or $\binom{3\cdot N}{3\cdot N-5}$ with filled cells for states $\ket{\Psi_d^{N}}$.
The conventional procedure of generating such types of states allows the randomly filled background to be distributed over the whole real space, i.e, over $N$ instead of $N'$ cells.
The purpose of our choice of initial states is to make participation ratios [see below] comparable for varying system sizes.

Starting with these configurations, we calculate the time evolution numerically and consider three different quantities which provide information on localization over time \cite{2017Santos,2020Santos}.
The results in Fig.~\ref{fig:time}(b) show the expectation values of the inhomogeneity operator (\ref{eq:A}) measured for $\ket{\Psi_{\text{s},\text{d}}^{10}(t)}$.
The initial value $\braket{A(0)}$ has to be equal for same initial states, however, the time evolution is strongly influenced by the value of $t_3$. 
For $t_3\neq 0$ the expectation values approach $\braket{A}_\text{eq}=1/2$ corresponding to thermal equilibrium and only vary weakly from then on.
If the band is flat for $t_3=0$, however, the relaxation values stay at $\braket{A}_\infty\approx 0.784$ for the sparse and $\braket{A}_\infty\approx 0.441$ for the dense system, showing that the system retains a memory of the initial state.

In Fig.~\ref{fig:time}(c) we show the revival probabilities $R(t)=\left|\braket{\Psi_{\text{s},\text{d}}^{10}(t)|\Psi_{\text{s},\text{d}}^{10}}\right|^2$ which give the probabilities for evolving states to return back into their initial states.
In the flat-band case the revival probability approaches a much larger value $R_\infty$ for long times around which $R(t)$ fluctuates.
The ratio between $R_\infty$ for $t_3=0$ and $t_3=1/4$ is $27.59$ in the sparse and $37.36$ in dense case.
Hence, it is in both cases many times more probable to return if the FB is present, again showing a higher degree of localization.

In Fig.~\ref{fig:time}(d) the time evolution of the participation ratio (PR) is shown.
It is defined as \cite{2015Beugeling,2017Santos}:
\begin{equation}
	PR(t)=\left[d\sum_{i=1}^{d}\left|\braket{\Psi_{\text{s},\text{d}}^{10}(t)|\chi_i}\right|^4\right]^{-1}\ ,
	\label{eq:pr}
\end{equation}
with $\ket{\chi_i}$ being the real-space many-particle basis.
\begin{figure}
	\centering
	\includegraphics[width=\linewidth]{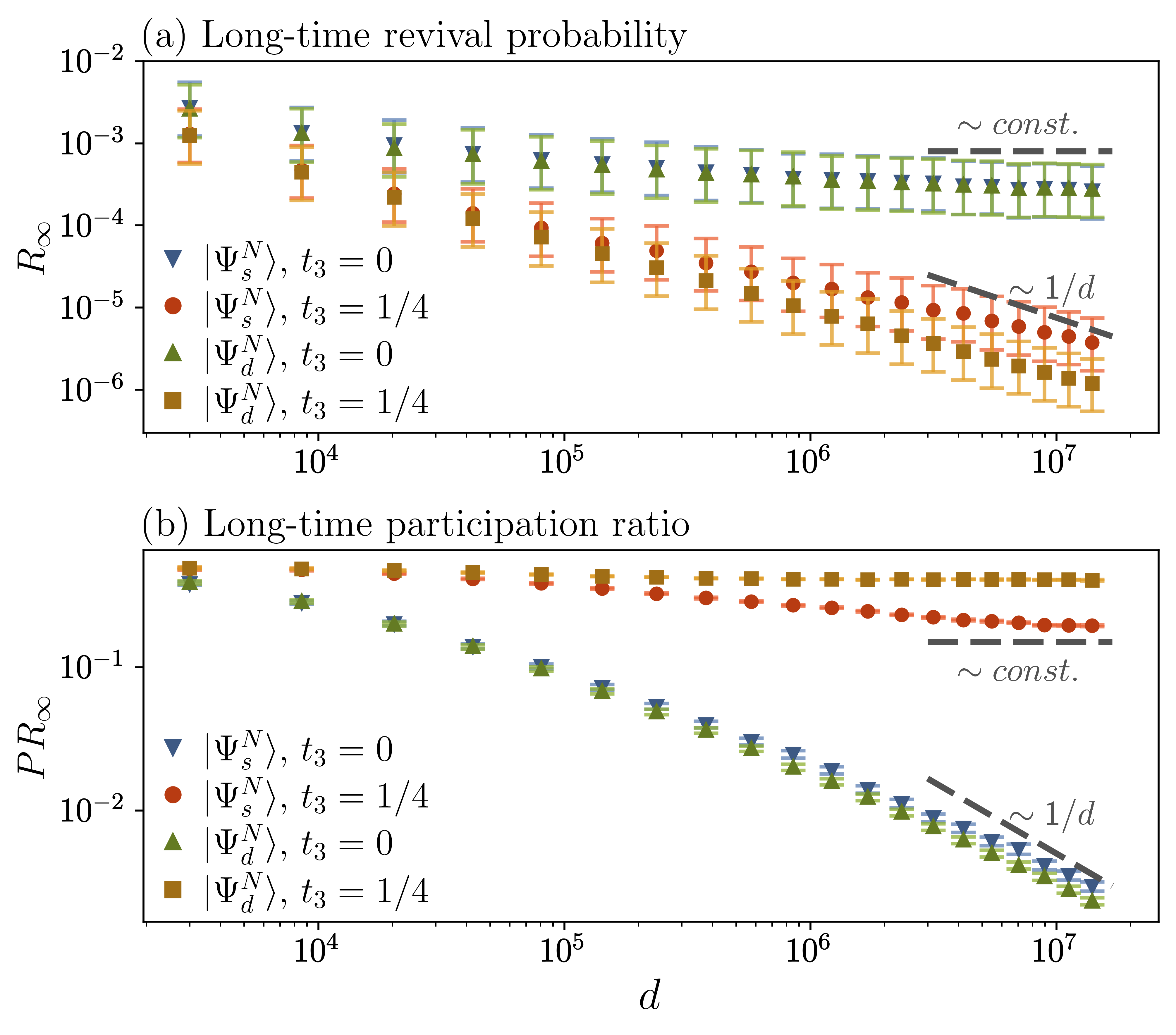}
	\caption{Long-time values $R_\infty$ (a) and $PR_\infty$ (b) vs. Hilbert space dimension $d$.
	The system sizes are from 12 to 60 lattice sites with either 5 particles or 5 holes. 
	The Hilbert-space dimension varies between $d=792$ and $d=5461512$.
	Data originate from single initial states.
	Scaling reference lines are shown for the largest dimensions.
    Error bars show the amount of fluctuation around the long-time value.}
	\label{fig:r_pr_dim}
\end{figure}
The participation ratio is a measure of how strongly a many-particle state is distributed over the real-space basis.
Possible values range from $1/d$ for a maximally localized state to $1$ for full delocalization.
Again, we can observe that the states in the system with FB tend to be significantly more localized at long times than without FB.
Ratios of the long-time participation ratios $PR_\infty$ between  $t_3=1/4$ and $t_3=0$ are $22.20$ (sparse) and $28.34$ (dense).\\

To judge more clearly the localization, it is of interest to study the dependence of the long-time values $R_\infty$ and $PR_\infty$ on the system size. 
To do so, we calculate them by removing or adding more unit cells to the system.
The initial states $\ket{\Psi_{\text{s},\text{d}}^{N}}$ are generalized for chains of different length.
The smallest systems possess $N=N'=5$ cells and therefore only contain the central region shown as blue area in Fig.~\ref{fig:time}(a).
Then, we enlarge the system by adding empty (sparse) or filled (dense) cells step by step up to $N=24$ with either $M=5$ or $M=3\cdot N-5$ particles.

In the limit of large chains, $R_\infty$ should be independent of the Hilbert-space dimension $d$ if the system is many-body localized and proportional to $d^{-1}$ if it is not.
Due to the factor $d$ in (\ref{eq:pr}), the opposite holds for $PR_\infty$.
The results in Fig.~\ref{fig:r_pr_dim} show that for large Hilbert-space dimensions in the FB case ($t_3=0$) $R_\infty$ approaches some constant value, while $PR_\infty$ adopts a $d^{-1}$ behavior, which one expects for a many-body localized system.
In contrast, for $t_3=1/4$ the revival probability $R_\infty$ decreases similar to $d^{-1}$ and $PR_\infty$ becomes constant, indicative for an ergodic system.
\section{Entanglement Entropy}
\label{sec:ee}
As a last test of MBL in a FB system we study the time dependent behavior of the bipartite von Neumann entanglement entropy (EE) in our diamond lattice.
\begin{figure}
	\centering
	\includegraphics[width=\linewidth]{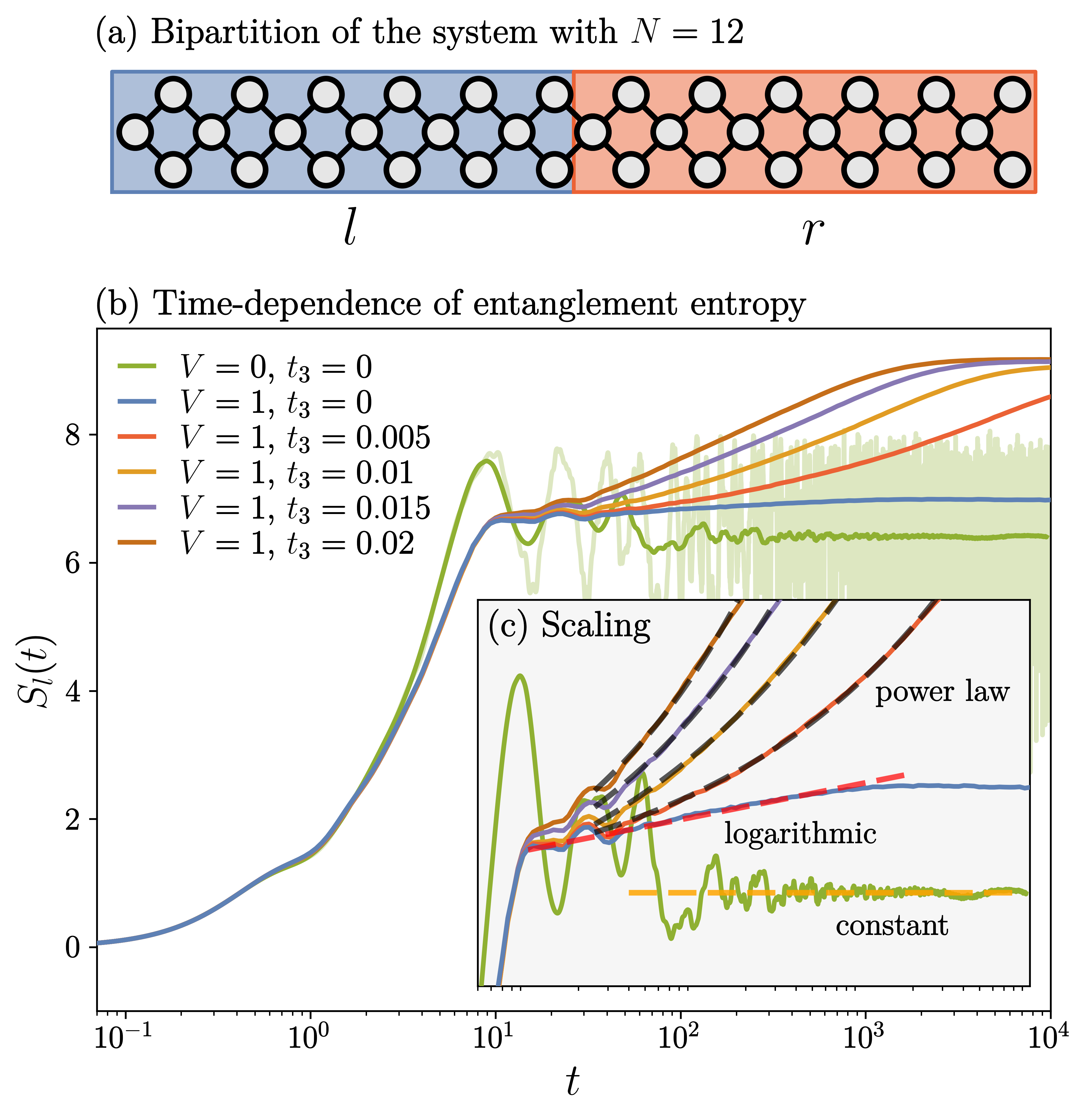}
	\caption{Time dependent EE data for a sparse system with filling $1/6$ and $N=12$ cells.
	(a) A sketch of how the system is split into left $l$ and right $r$ subsystems.
	(b) Plot of EE results stem from averaging over a few random detangled initial states. 
	The noisy data of the interactionless case $V=0$ has been smoothed with Savitzky–Golay filtering \cite{1964Savitzky}.
	The inset (c) presents a zoom in on the different scaling behaviors which are difficult to detect at full scale.
	For $V=0$ the EE becomes constant while a slow logarithmic incline is visible for $V=1$. 
	As soon as the FB is slightly perturbed by $t_3\neq0$ a power law scaling sets in.}
	\label{fig:ee_time}
\end{figure}
It is defined as\cite{2006Popescu,2013Bauer,2014Grover,2015Beugeling,2018Mori,2019Sirker,2021Orito}:
\begin{equation}
	S_l=-\operatorname{tr}\left[\rho_l\log_2(\rho_l)\right]\ ,
	\label{eq:ee}
\end{equation}
with $\rho_l$ being the reduced density matrix of the left subsystem $l$ [see Fig.~\ref{fig:ee_time}(a)].
$S_l$ is measured in a time evolution of non-entangled product states which are used as initial state configuration $\ket{\Psi}$.
They are constructed with either subsystem $l$ or $r$ [Fig.~\ref{fig:ee_time}(a)] being filled resp. empty:
\begin{equation}
	\ket{\Psi}=\ket{0}_l\otimes\ket{\Phi}_r\quad\text{or}\quad\ket{\Psi}=\ket{\Phi}_l\otimes\ket{0}_r\ ,
	\label{eq:ee_is}
\end{equation}
where $\ket{\Phi}$ is a random state with complex Gaussian distributed coefficients (mean zero).
	
Referring to Table 1 in \cite{2015Nandkishore} a canonical classification of EE with time can be issued as follows.
In a single-particle localized system we should not observe any spreading of EE after some initial incline. 
A slow logarithmic increase can be observed in the MBL phase while the thermal phase reveals itself by a faster power law increase.
In Fig.~\ref{fig:ee_time}(b) and (c) we show that our results are consistent with this categorization.
Independent of the specific set of parameters a rather similar short-lived incline sets in from initial state expansion after the time evolution has been started.
From approximately $t\approx10$ on the subsequent behavior is determined by the value of interaction strength $V$ and perturbation of the FB $t_3$.
In the interactionless case with perfect FB no further incline can be found after application of noise filtering on the data which is consistent with single-particle localization.
Presence of interaction leads to much less noise and a slow increase of the EE consistent with a logarithmic increase expected in MBL systems. 
Perturbation of the FB causes a power law scaling $\sim t^{\alpha}$ of $S_l(t)$ until the EE reaches a finite size limit.
	
Another property regarding entanglement is area law vs. volume law scaling with system size.
This matter is discussed in Appendix B.
\section{Conclusion}
\label{sec:concl}
To summarize, we examined quasi one-dimensional interacting diamond ladders
which provide FBs.
The configurations were chosen to model four cases: sparse filling with and without FB and dense filling with and without FB.
The flatness was controlled via an inter-cell hopping perturbation.
We have shown that the FB cases violate the eigenstate thermalization hypothesis. 
When considering FB systems the degree of degeneracy is a key element for localization.
An exact calculation demonstrated that degeneracy is not a phenomenon of finite system sizes but grows extensively in the thermodynamic limit.
This is likewise true when interaction is added.
Time dependence of selected nonequilibrium initial states showed that the system retains a memory of the initial state.
\begin{figure}
	\centering
	\includegraphics[width=\linewidth]{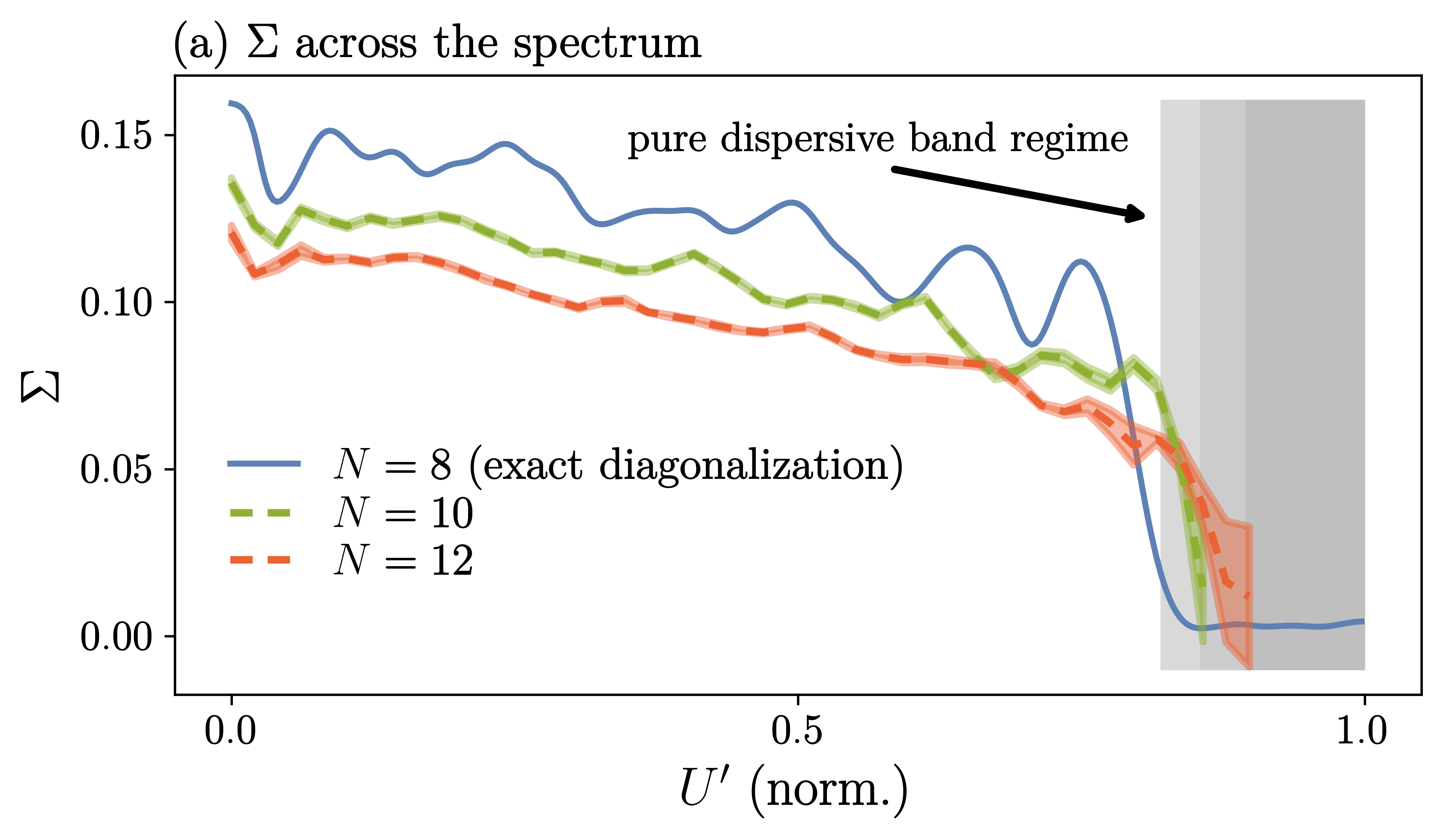}
	\caption{Operator expectation value fluctuations $\Sigma$ for the sparse system with filling $1/6$ with $V=1$ and $t_3=0$.
	$N=8$ has been calculated with exact diagonalization while for $N=10,12$ the typicality scheme described in section~\ref{sec:eth} is used.
	The width of the energy filter has been set to a constant $\sigma=1/2$ which is $\approx1.6\%$ ($N=8$), $\approx1.3\%$ ($N=10$) and $\approx1.1\%$ ($N=12$) of the spectrum.
	The energy range has been normalized between $0$ and $1$ with variable $U'$ as filter.
	Setting $U'$ to $0$ corresponds to filtering at the groundstate energy while $U'=1$ means filtering at the energy of the highest excited state.
	$\Sigma$ only vanishes in the highest energy range with vanishingly low density of states.
	Data points near $U'=1$ are omitted due to lack of statistics.
	A shaded confidence interval around $\Sigma$ gives the standard deviation from using different random initial states.}
	\label{fig:app_eth}
\end{figure} 
Finite size scaling of the revival probability and participation ratio was shown to be consistent with localization. 
Time dependence of the EE revealed behavior consistent with slow logarithmic spreading in the FB MBL case, while it remained constant in the single-particle localized case.
As soon as the FB is perturbed a power law scaling could be found.
Taken together, our numerical results show that several key characteristics of MBL are obeyed by the quasi one-dimensional interacting diamond ladders.
This system thus provides an example for MBL in a defect-free system.
\section*{Acknowledgments}
Financial support from the DFG via the research group FOR2692, grant numbers
397171440 and 397067869 is gratefully acknowledged. We would like to thank S. kleine Br\"uning and S. Tilleke for valuable discussions.
\section*{Appendix A: Violation of ETH in Excited States}
\label{sec:app_eth}
As mentioned above it is possible to expand ETH calculations to any energy regime with the described scheme.
Instead of setting the energy filter $U$ in (\ref{eq:filter}) to the groundstate, it can be varied from the lowest to the highest energy in the system.
If localization caused by a FB is supposed to be a robust mechanism for MBL the FB should also impact higher excited states.

The results are shown in Fig.~\ref{fig:app_eth} on a normalized energy scale.
$\Sigma$ only vanishes for the highest energies which - in the sparse system - can be identified as regime of the dispersive band. 
This energy range is governed only by extended dispersive band states.
Its width and participation in the spectrum narrows with increasing system size which supports the reasoning concerning extensive degeneracy in Fig.~\ref{fig:degen}.
\section*{Appendix B: Area and Volume Law of the Entanglement Entropy}
\label{sec:app_ee}
In contrast to the time dependent spreading characteristics of entanglement in Fig.~\ref{fig:ee_time} where clear distinctions between MBL and the thermal phase could be found, we are only able to demonstrate trends when it comes to area law vs. volume laws in our model.
\begin{figure}
	\centering
	\includegraphics[width=\linewidth]{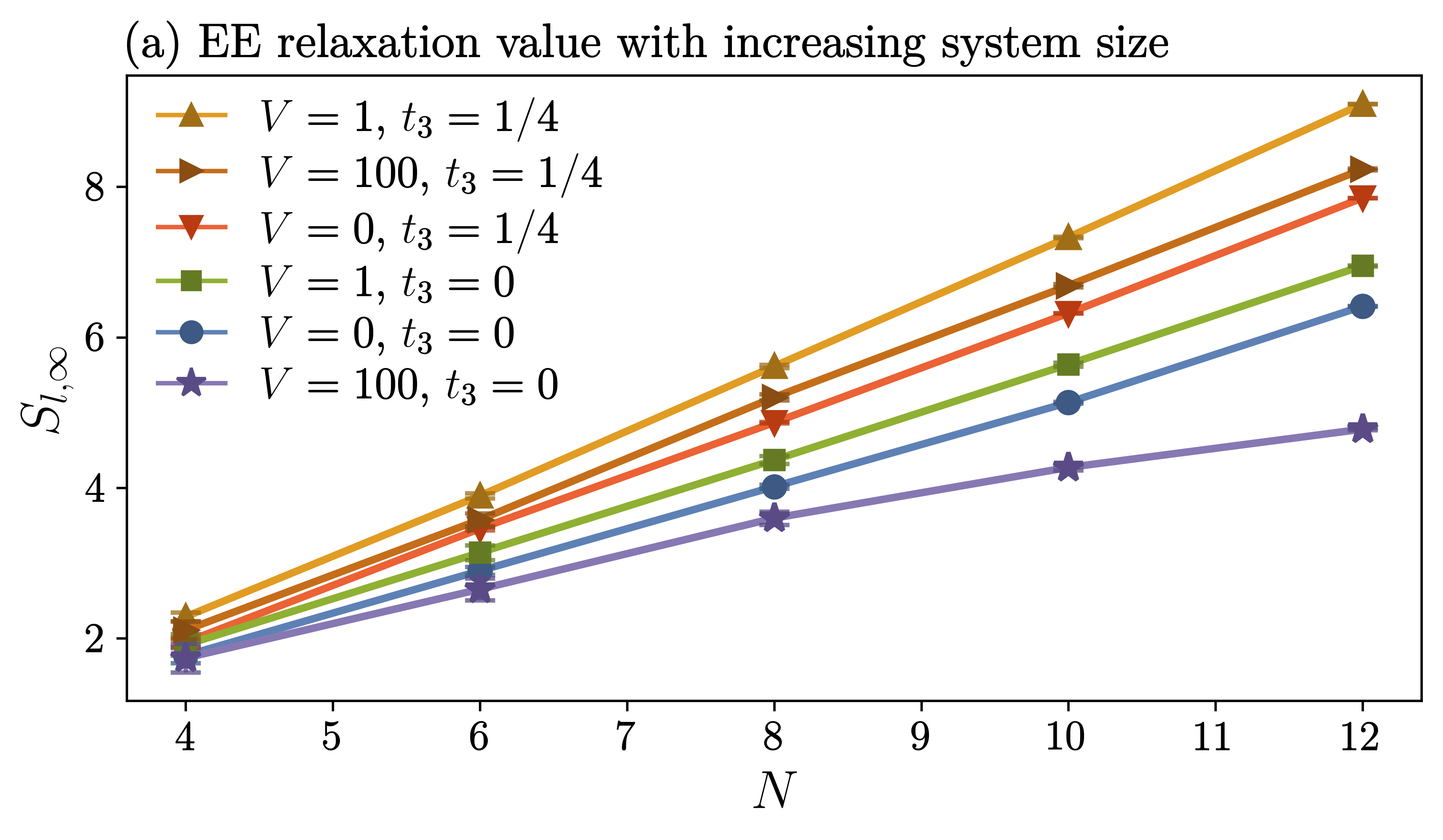}
	\caption{EE data for increasing system size.
	The system is sparsely filled with $1/6$. $S_{l,\infty}$ is calculated as long-time relaxation value and averaged over many random initial states (\ref{eq:ee_is}).
	Error bars due to typicality averaging are depicted yet too small to be visible.}
	\label{fig:app_ee}
\end{figure}
Based again on Table 1 in \cite{2015Nandkishore} area law is expected in the single-particle localized and in the MBL phase.
In a one-dimensional system this corresponds to EE becoming constant with increasing system size.
For a thermal system EE adopts a volume law which means linear scaling.
We calculate a long-time relaxation value $S_{l,\infty}$ with (\ref{eq:ee}) similar to Fig.~\ref{fig:r_pr_dim} to address this question.
The initial states are again detangled product states (\ref{eq:ee_is}) with a filled and an empty partition of the system.

Results are presented in Fig.~\ref{fig:app_ee}.
A distinct volume law scaling can be found in every scenario with perturbation $t_3=1/4$, independent of interaction strength $V$.
When it comes to the single particle localized scenario with $V=0$ [blue curve] and moderate interaction $V=1$ [green curve] still no clear deviation from volume law scaling can be observed.
Contribution of extended modes from dispersive bands with localization lengths larger than the examined system sizes prevent containment of spatial entanglement at these small scales. 
The localization length can be reduced by increasing the interaction and thus reinforcing the coupling between extended and FB modes.
Static FB states then act as a form of annealed disorder for dispersive states \cite{2022Danieli}.
We see that for strong interaction $V=100$ [purple curve] an approach to area law scaling can be surmised.

\bibliographystyle{apsrev4-1}
\bibliography{mbl_tidl_fb_v5}

\end{document}